# MICROLENSING ANALYSIS FOR THE GRAVITATIONAL LENS SYSTEMS SDSS0924+0219, Q1355-2257, AND SDSS1029+2623

K. Rojas,[1,2] V. Motta,[2] E. Mediavilla,[3,4] J. Jiménez-Vicente,[5,6] E. Falco,[7] and C. Fian[3,4]

[1]*Institute of Physics, Laboratoire d'Astrophysique, Ecole Polytechnique Fédérale de Lausanne (EPFL), Observatoire de Sauverny, CH-1290 Versoix, Switzerland*
[2]*Instituto de Física y Astronomía, Universidad de Valparaíso, Avda. Gran Bretaña 1111, Valparaíso, Chile*
[3]*Instituto de Astrofísica de Canarias, Avda. Vía Láctea s/n, La Laguna, Tenerife 38200, Spain*
[4]*Departamento de Astrofísica de Canarias, Universidad de La Laguna, Avda. Vía Láctea s/n, La Laguna, Tenerife 38200, Spain*
[5]*Departamento de Física Teórica y del Cosmos, Universidad de Granada, Campus de Fuentenueva, E-18071 Granada, Spain*
[6]*Instituto Carlos I de Física Teórica y Computacional, Universidad de Granada, E-18071 Granada, Spain*
[7]*Whipple Observatory, Smithsonian Institution, 670 Mt. Hopkins Road, PO Box 6369, Amado, AZ 85645, USA*

## ABSTRACT

We use spectroscopic observations of the gravitationally lensed systems SDSS0924+0219(BC), Q1355-2257(AB), and SDSS1029+2623(BC) to analyze microlensing and dust extinction in the observed components. We detect chromatic microlensing effects in the continuum and microlensing in the broad emission line profiles of the systems SDSS0924+0219(BC), and Q1355-2257(AB). Using magnification maps to simulate microlensing and modeling the emitting region as a Gaussian intensity profile with size $r_s \propto \lambda^p$, we obtain the probability density functions for a logarithmic size prior at $\lambda_{rest-frame}$ = 3533 Å. In the case of SDSS0924+0219, we obtain: $r_s = 4^{+3}_{-2} \sqrt{M/M_\odot}$ light-days (at $1\sigma$), which is larger than the range of other estimates, and $p = 0.8 \pm 0.2$ (at $1\sigma$), which is smaller than predicted by the thin disk theory, but still in agreement with previous results. In the case of Q1355-2257 we obtain (at $1\sigma$): $r_s = 3.6^{+3.0}_{-1.6} \sqrt{M/M_\odot}$ light-days, which is also larger than the theoretical prediction, and $p = 2.0 \pm 0.7$ that is in agreement with the theory within errors. SDSS1029+2326 spectra show evidence of extinction, probably produced by a galaxy in the vicinity of image C. Fitting an extinction curve to the data we estimate $\Delta E \sim 0.2$ in agreement with previous results. We found no evidence of microlensing for this system.

## 1. INTRODUCTION

The study of the inner structure of distant quasars (QSOs) is a big challenge, because this region is small and it is not resolvable with current observational facilities. As a consequence, theories of accretion disks, which predicted accretion disk sizes and their radial temperature profile (Shakura & Sunyaev 1973), still need to be proven. Indirect observational evidence, such as reverberation mapping (Wanders et al. 1997; Collier et al. 1998; Edelson et al. 2015) or microlensing (Moustakas & Metcalf 2003; Anguita et al. 2008; Bate et al. 2008; Floyd et al. 2009; Sluse et al. 2012; Motta et al. 2012; Guerras et al. 2013; Jiménez-Vicente et al. 2014; Rojas et al. 2014; Mediavilla et al. 2015; Motta et al. 2017), have found that accretion disks are larger than predicted by theory. Both techniques require long monitoring campaigns in multiple wavelength bands, but microlensing can also be studied using single-epoch spectra, which is an advantage with regard to observing time.

Microlensing by stars in the lens galaxy causes flux variations in QSO lensed images (Chang & Refsdal 1979; Wambsganss 2006). This effect is size sensitive, producing large magnifications of sources with angular size comparable to (or smaller than) the microlens Einstein radius. The flux of larger regions, like the Narrow Line Region (NLR), is expected to be insensitive to microlensing, while inner regions, such as the accretion disk and the inner parts of the Broad Line Region (BLR), can be affected by microlensing. Theoretical work by Shakura & Sunyaev (1973) shows that the size of the accretion disk ($r_s$) varies with wavelength ($r_s \propto \lambda^p$). This variation produces a chromatic microlensing effect (Wambsganss & Paczynski 1991; Wisotzki et al. 1995; Mosquera et al. 2009; Mediavilla et al. 2011) because the magnification will be different depending on the wavelength (region of the disk), i.e. the flux variation is stronger for shorter wavelengths and almost negligible in the infrared (IR).

However, there are other reasons for chromatic variations such as QSO intrinsic variability coupled with time delays between images and dust extinction. Following Yonehara et al. (2008), for a pair of images with small separation (<2.0"), a time delay of ∼ 40 days produces a chromaticity change <0.03 mag. Then, the effect of chromaticity due to intrinsic variability is negligible for the objects presented in this work. On the other hand, dust extinction affects both continuum and emission lines. Then, the chromatic variation trend will be seen in all measurements. For these reasons, special care is required to distinguish among microlensing, chromaticity, and differential dust extinction along the path towards each image through the lens galaxy.

In this paper we present single-epoch spectra for the lensed quasars SDSS0924+0219, Q1355-2257, and SDSS1029+2623 to search for perturbations produced by microlensing, chromatic microlensing and/or dust extinction. In the cases of chromatic microlensing detection, the measurements will be used to improve the statistical analysis of quasar accretion disk sizes as shown in previous studies (Jiménez-Vicente et al. 2014, 2015a,b). In Section 2 we present the data and reduction technique used. We explain the methods applied to perform the different analysis in Section 3. In Section 4 we discuss our results, and the conclusions are presented in Section 5.

## 2. OBSERVATIONS AND DATA REDUCTION

We obtained VLT/FORS2 spectra for SDSS0924+0219 and Q1355-2257 in 2008/04/02 (P.I. V. Motta, 381.A-0508), with a seeing of 0.8". We used the grism 300V, with resolution of 11.0 Å pixel$^{-1}$, and the blocking filter GG435. The wavelength range of the spectra is 2946-9645 Å. The position angle of the slit (in degrees E to N) was chosen to observe simultaneously two images of the quasar: $42^o$ to observe C and B for SDSS0924+0219, and $-72^o$ to observe A and B for Q1355-2257. The exposure times are: $10 \times 1800$ s for SDSS0924+0219, and $3 \times 540$ s for Q1355-2257, with an average airmass of 1.1". In the case of Q1355-2257, we additionally used the deconvolved spectra from Sluse et al. (2012) provided by the VizieR[1] catalog (Ochsenbein et al. 2000).

The data reduction was performed with IRAF tasks and included bias subtraction, flat normalization, and wavelength calibration. The flux calibration was not applied since we are only interested in flux ratios. The spectra extraction, in the case of Q1355-2257, was made by fitting a Gaussian to each component through each wavelength bin, because the separation between the components is 1.03" and the spectra overlap. In the case of SDSS 0924+0219, the separation is also small (1.18") and part of the flux of the lens galaxy is contaminating the quasar components. In this case we fitted three Gaussians: two for the images and one for the lens galaxy. To better constrain the lens spectrum we guided the lens galaxy flux amplitude using a simulated spectrum of the galaxy. As the spectra were taken with FORS, we model the lens galaxy spectrum using the ESO Exposure Time Calculators[2] for FORS2 instrument. We choose the template of an elliptical galaxy spectrum (Kochanek et al. 1999) at redshift 0.39 with magnitude V=21 mag. The sky conditions and instrument setup were selected to match those of our observation. Figures showing the wavelength calibrated and extracted spectra are in Appendix A

---

[1] Based on data obtained with the VizieR catalog access tool, CDS, Strasbourg, France.
[2] https://www.eso.org/observing/etc/

For both systems, we also used data from CASTLES[3]. These were taken in 2003 with the Hubble Space Telescope (HST) in three different bands (F160W, F555W, and F814W). Additionally, we used data from the literature for SDSS0924+0219 (Inada et al. 2003; Pooley et al. 2007; Floyd et al. 2009; Blackburne et al. 2011), and Q1355-2257 (Morgan et al. 2003; Sluse et al. 2012).

SDSS J1029+2623 spectra were taken with the LRIS-ADC at Keck and reduced spectra were kindly provided by M. Oguri. The details related to the observation and data reduction are explained in Oguri et al. (2008).

## 3. METHOD

To distinguish between microlensing, chromatic microlensing and extinction we compare the magnitude difference in the continuum under the emission lines with the magnitude difference in the emission line cores, $\Delta m = (m_B - m_A)_{cont} - (m_B - m_A)_{core}$ (e.g., see Mediavilla et al. 2009, 2011; Rojas et al. 2014; Motta et al. 2017). We perform this analysis using a set of python packages[4]. The continuum is obtained by fitting a straight line to regions selected at both sides of each emission line. Then, the continuum is subtracted, and we integrate the emission line in a relatively narrow interval (between 20 to 60 Å, depending on the line shape) around the peak (hereafter: core of the line). For those emission lines with absorption around the core (e.g. CIV in SDSS1029+2623) a much narrower interval was selected (20-25 Å). The uncertainty in the continuum is the fit root mean square (rms) error, and for the lines it is the rms error in the determination of the total flux added in quadrature, which is assumed to be the same as the continuum.

If $(m_B - m_A)_{core}$ does not present any change with wavelength, we use the mean of all values as a no-microlensing baseline. Thus, if $(m_B - m_A)_{cont} \neq (m_B - m_A)_{core}$ (i.e. $\Delta m \neq 0$) it means there is microlensing. Furthermore, if these $|\Delta m|$ values change with wavelength, being larger at blue than at red wavelengths, tentatively the system exhibits chromatic microlensing.

For each pair of images, we compare the profiles of the emission lines after the continuum substraction to search for microlensing in the Broad Line Region (BLR). If we find differences between the profiles, we integrate the line using windows of 25-30 Å (in the rest frame) for both the red and blue wings. The magnitude difference between the lines is used to estimate the size of the BLR (Guerras et al. 2013; Motta et al. 2017).

For those systems with no lens model available in the literature, we calculate our own by using a Singular Isothermal Ellipsoid (SIE) in *Lensmodel* (Keeton 2001). We employ the astrometry available in CASTLES and the measured flux ratios of the emission line cores. The model provides the convergence and shear at the position of each image ($\kappa_A$, $\gamma_A$, $\kappa_B$, $\gamma_B$), which are used to compute magnification maps applying the Inverse Polygon Mapping method (Mediavilla et al. 2006, 2011). We considered microlenses of 1 $M_\odot$ and assumed a mass fraction in stars $\alpha = 0.1$ (Mediavilla et al. 2009; Pooley et al. 2009). The size for each map is 15×15 Einstein radii (5000×5000 pixels$^2$).

To estimate the size of the accretion disk and its temperature profile from the microlensing data, we follow a Bayesian procedure (see, e.g. Mediavilla et al. 2011). Mortonson et al. (2005) have shown that the magnification statistics of microlensing are dependent on the half light radius of the source, but are quite insensitive to the specific radial profile of the source. We have used a Gaussian intensity profile I(R) $\propto$ exp(-$R^2/2r_s^2$), where $r_s$ is the accretion disk size and p is related to the temperature profile of the disk (p=1/$\beta$). This profile is computationally convenient, for which $r_{1/2}$=1.18$r_s$ (being $r_s$ the sigma of the Gaussian). For a thin disk, (Shakura & Sunyaev 1973), p=4/3. To estimate the likelihood of reproducing the measured microlensing amplitudes we randomly place Gaussian sources with different sizes and profile slopes on the magnification maps, using a logarithmic prior on $r_s$. All the estimations were obtained at $\lambda_{rest-frame}$ = 3533Å.

We expect that the cores of the emission lines are not affected by microlensing, but we could find differences in $(m_B - m_A)_{core}$ with wavelength produced by extinction from different amounts of dust and gas in the lens galaxy. Using the Cardelli et al. (1989) extinction law, we fit the data using the equation: $m_1(\lambda) - m_2(\lambda) = -2.5\log(M_1/M_2) + (\Delta E)R_v(\lambda/1 + z_L)$ (Falco et al. 1999), where $\Delta M_0 = M_1/M_2$ is the magnification ratio, $\Delta E = E_1 - E_2$ is the differential extinction and $R_v$ is the parametrized extinction law. We considered two cases: one where we left $R_v$ as a free parameter, and a second one with a fixed $R_v$ = 3.1 corresponding to the Milky Way extinction curve.

## 4. RESULTS

### 4.1. *SDSS0924+0219*

This quadruple system was discovered by Inada et al. (2003) who found a quasar redshift of $z_S$=1.52. The lens is an elliptical galaxy at $z_L$=0.39 (Ofek et al. 2006). We present VLT spectra (see Figure 1) for the C and B components which are separated by 1.52". The line profiles show differences in the line wings likely produced by microlensing in the red wing of CIII], but a single measurement is not enough to study the BLR structure. The blue wings of CIII] and MgII and the red wing of MgII are probably affected by underlying lines like FeII.

---

[3] CFA-Arizona Space Telescope LEns Survey, Kochanek, C.S., Falco, E., Impey, C., McLeod, B., & Rix, H.W. http://www.cfa.harvard.edu/glensdata.
[4] https://github.com/Krojas/QSO_microlensing

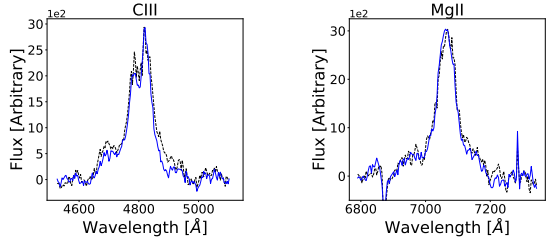

**Figure 1.** CIII] and MgII emission lines profiles without the continuum for SDSS0924+0219. In black the A component and in blue the B component multiplied by a factor of 1.38 (CIII]) and 1.33 (MgII) respectively.

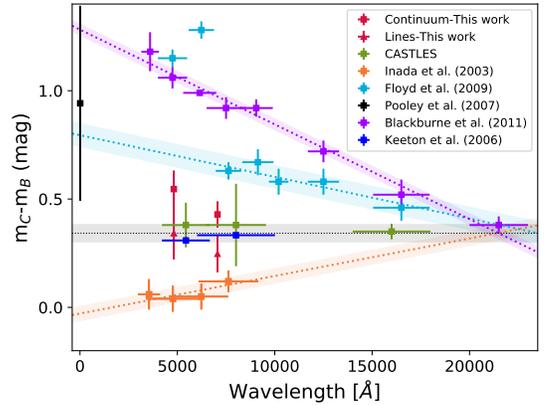

**Figure 2.** Magnitude differences $m_C$-$m_B$ as a function of wavelength for SDSS0924+0219. Red triangles are the emission line cores and the red squares are the continuum calculated from VLT spectra. Orange squares are the photometric data presented by Inada et al. (2003) for the bands: ugri. Green squares are the broad band continuum from CASTLES. Blue squares are Keeton et al. (2006) measurements for HST bands F555W and F816W. Purple squares are the relative photometry for the bands u'g'r'i'z'JHK$_s$ in Blackburne et al. (2011). The black square is an X-ray measurement in the 0.5-8 keV band given by (Pooley et al. 2007). Light blue squares are photometric data of Floyd et al. (2009) for the bands: HJYz'i'r'g'. The black dotted line is the median among the measurements of the line cores, CASTLES data and Keeton et al. (2006) bands, and the shaded area around is the linear fit standard deviation. The colored dotted lines represent the best linear fit for each set of points, the color shaded region represent the error in the linear fit associated with magnitude difference.

We calculate the magnitude differences for the cores of the emission lines and the continua ($\Delta m$) for the spectra (Figure 2 and Table 2) and compared them with values from the literature. CASTLES and Keeton et al. (2006) values are in agreement with our measurements and with the IR values, meaning that there is no microlensing effect in that epoch (2003). We take as no-microlensing baseline the median value of the cores of the lines, HST broad-band, and Keeton et al. (2006) data: $0.34 \pm 0.04$ mag. In our data we find $\Delta m$ up to 0.34 mag at $\lambda 4800$ but it decreases to $\sim 0.25$ mag at $\lambda 7000$. We decided against chromatic microlensing analysis of this data set because there is no clear evidence for this effect. A clearer chromatic trend can be seen in the values taken from literature.

The first data set correspond to Inada et al. (2003), taken in 2001. The HST data (CASTLES and Keeton et al. (2006)) were taken in 2003 and they all agree with the measurements of the cores of the emission lines, indicating that the continuum of each component is not affected by microlensing. Comparing Blackburne et al. (2011) (data taken between February 2007 and May 2008), and Floyd et al. (2009) (observed in March 2008), we see a change in the slope of the chromatic microlensing. The points corresponding to the g and r bands in Floyd et al. (2009) show large differences compared to the other bands. Considering that the authors do not offer an explanation for this, we are unable to re-analyze the data to confirm those measurements (Magellan telescopes have no public data archive), and that Blackburne et al. (2011) data show no such deviations (around the same epochs), we decided to disregard them in our calculations.

The differences between the baseline and the continua obtained from different data sets are listed in Table 3. We analyze the three continuum data sets (Inada et al. 2003; Blackburne et al. 2011; Floyd et al. 2009) independently. We also estimate the differences between the CIII] red wing and the core of the emission lines obtaining $0.63 \pm 0.17$ mag.

As the probability density function for each data set is an independent measurement of $r_s$ and p, we calculate their product (see Figure 3). This distribution gives a size for the accretion disk of $r_s = 4^{+3}_{-2} \sqrt{M/M_\odot}$ light-days at $\lambda 3533$ Å ($\ln((r_s/\text{light-days})$

$$\sqrt{M/M_\odot})$$

$= 1.5 \pm 0.5$) and a value for the thermal profile $p = 0.8 \pm 0.2$ at $1\sigma$. From the thin disk theory, we expect $r_s = 0.12 \sqrt{M/M_\odot}$ light-days, assuming a black hole mass of $2.8 \times 10^8 M_\odot$ (Morgan et al. 2006), an Eddington ratio of $L/L_E = 0.1$ and a radiative efficiency of $\eta = 0.1$. This predicted size is significantly smaller than our measurement, but is in agreement with other estimates: $0.07 < r_{s,\lambda 2770} < 0.26$ $\sqrt{M/M_\odot}$ light-days (Morgan et al. 2006), $r_{s,\lambda 3200} = 0.14$ $\sqrt{M/M_\odot}$ light-days (Pooley et al. 2007), $r_{s,\lambda 2770} = 0.39^{+0.38}_{-0.24}$ $\sqrt{M/M_\odot}$ light-days (Morgan et al. 2010), $r_{s,\lambda 8140} = 0.44$ light-days $\sqrt{M/M_\odot}$ (Mosquera & Kochanek 2011), $r_{s,\lambda 2500} = 0.61^{+0.4}_{-0.3}$ $\sqrt{M/M_\odot}$ light-days (MacLeod et al. 2015). On the other hand Blackburne et al. (2011) calculated $r_{s,\lambda 3462} = 1.3^{+0.6}_{-0.4}$ $\sqrt{M/M_\odot}$ light-days, and Floyd et al. (2009) $r_{s,\lambda 3540} = 4.7$ $\sqrt{M/M_\odot}$ light-days (for all data presented in) which are in agreement with our estimation. With respect to the temperature index Floyd et al. (2009) estimate p=0.75 using only Magellan data, and $0.8 < p < 17$ using all their data set, and MacLeod et al. (2015) estimated $p = 2.17 \pm 2.17$. Our result is in agreement with Floyd et al. (2009) and also with the average slope found by Jiménez-Vicente et al. (2014) (p = 0.8) using a sample of 8 systems.

**Table 2.** SDSS0924+0219 Magnitude differences

| Region | $\lambda$ (Å) | Window[a] (Å) | $m_C$-$m_B$ (mag) |
|---|---|---|---|
| Continuum | 4818 | 4543-5174 | $0.55 \pm 0.09$ |
|  | 7067 | 6590-7319 | $0.43 \pm 0.06$ |
| Line | CIII] | 4805-4830 | $0.34 \pm 0.12$ |
|  | CIII] red wing | 4853-4904 | $0.97 \pm 0.12$ |
|  | MgII | 7049-7075 | $0.25 \pm 0.09$ |

[a] Integration window.

Table 3. SDSS0924+0219 Chromatic Microlensing

| Data | λ (Å) | Δm (mag) |
|---|---|---|
| Inada et al. (2003) | 3545.0 | $-0.31 \pm 0.06$ |
|  | 6231.0 | $-0.26 \pm 0.06$ |
|  | 7625.0 | $-0.24 \pm 0.06$ |
| Blackburne et al. (2011) | 3600.0 | $0.78 \pm 0.05$ |
|  | 9050.0 | $0.54 \pm 0.06$ |
|  | 21500.0 | $-0.00 \pm 0.06$ |
| Floyd et al. (2009) | 7625.0 | $0.31 \pm 0.10$ |
|  | 10200.0 | $0.26 \pm 0.09$ |
|  | 16500.0 | $0.14 \pm 0.09$ |

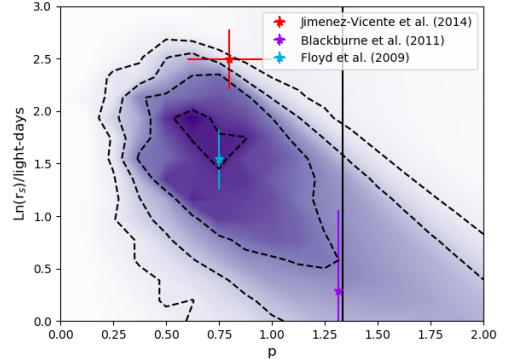

**Figure 3.** Combined probability density function for SDSS0924+0219 using a logarithmic size prior for the different $\Delta m$ values. Contours correspond to $0.5\sigma$, $1\sigma$, $1.5\sigma$, and $2\sigma$ respectively. The black solid line shows the value for p predicted by Shakura & Sunyaev (1973). The light blue star represent the measurements presented by Floyd et al. (2009). The purple star show the accretion disk size measure by Blackburne et al. (2011), the value of the temperature profile have been shift -0.02 for visualization purposes. The red star represents the average size and temperature for the sample analyzed by Jiménez-Vicente et al. (2014).

### 4.2. Q1355-2257

Discovered by Morgan et al. (2003) using HST data, it is a double lensed quasar with an image separation of 1.23". The redshift of the quasar is $z_S$=1.37 (Morgan et al. 2003) and the redshift for the lens is $z_L$=0.70 (Eigenbrod et al. 2006).

We present VLT spectra for both images of the quasar (AB), and also a re-analysis of the deconvolved data (see Appendix A) presented by Sluse et al. (2012). The line profiles for both data sets (Figure 4) show small differences in the wings but are likely produced by microlensing.

The magnitude differences for the core of the emission lines and the continua below are shown in Figure 5 and Table 4. Comparing our results for the deconvolved data and the values published in Sluse et al. (2012) we notice a discrepancy between both results of -0.06 mag for the continuum and +0.08 mag for the lines. This could be explained due to differences in the method to analyze the data. In the mentioned publication, they fitted Gaussian profiles for the NLR and BLR while we integrate restrictive windows around the core of the emission line to avoid taking into account flux affected by microlensing coming from of the BLR. We obtained a no microlensing baseline of $1.24 \pm 0.04$ mag calcu-lating the median among the emission lines cores from our data, the results published by Sluse et al. (2012) and the one obtained with our method.

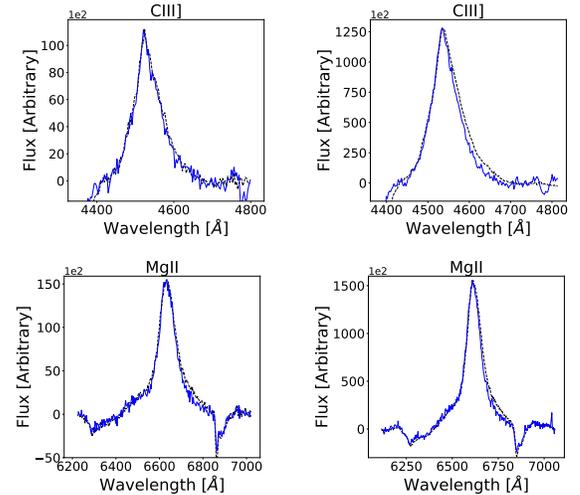

**Figure 4.** CIII] and MgII emission line profiles without the continuum for Q1355-2257. This work (left panel) and our re-analysis of Sluse et al. (2012) spectra (right panel). Shown in black is the A component and in blue is the B component multiplied by a factor of 3.3 (CIII]), and 3.1 (MgII) for our data and 3.3 (CIII]), and 3.2 (MgII] for deconvolved data.

In general, there is a difference of at least 0.3 mag between the baseline and continuum from our data, Sluse et al. (2012) deconvolved spectra, and Morgan et al. (2003) data, corroborating the presence of microlensing. We found evidence of chromatic microlensing for wavelengths greater than $\lambda > 6180$Å. The magnitude difference presents a change in the slope similar to the one seen in the F814W and F160W broad-bands.

Table 4. Q1355-2257 Magnitude differences

| Region | | $\lambda$ (Å) | Window[a] (Å) | $m_C$-$m_B$ (mag) |
|---|---|---|---|---|
| Continuum | | | | |
| | Our data | 4521 | 4400-4764 | $1.73 \pm 0.17$ |
| | | 6633 | 6050-7099 | $1.63 \pm 0.03$ |
| | Deconvolved data | 4531 | 4418-4764 | $1.79 \pm 0.35$ |
| | | 6608 | 6050-7099 | $1.61 \pm 0.02$ |
| Line | | | | |
| | Our data | CIII] | 4506-4536 | $1.28 \pm 0.24$ |
| | | MgII | 6603-6663 | $1.23 \pm 0.04$ |
| | Deconvolved data | CIII] | 4520-4551 | $1.30 \pm 0.49$ |
| | | MgII | 6585-6635 | $1.25 \pm 0.03$ |

[a] integration window.

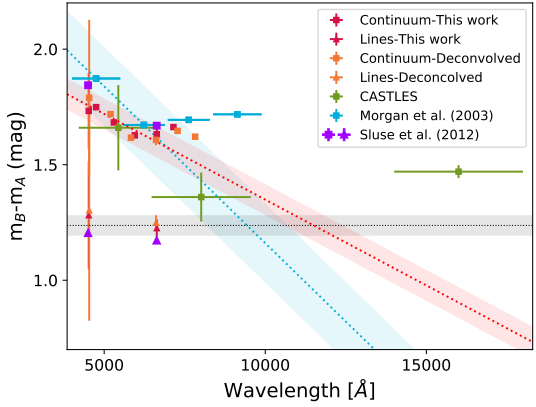

**Figure 5.** Magnitude differences $m_B$-$m_A$ as a function of wavelength for Q1355-2257. Triangles are the emission line cores without the continuum, and squares are the integrated continuum under the line cores. Red symbols are the measurements obtained from our VLT spectra. Orange symbols are the results obtained using the deconvolved data of Sluse et al. (2012), while the purple symbols are the estimation presented by the authors, with error bars smaller than the symbols. Green squares are the continua obtained from CASTLES. Light blue squares are the data for the g, r, i, and z bands in Morgan et al. (2003). The black dotted line is the median of the emission lines, and the grey shaded area represents the standard deviation. The colored dotted lines are the best linear fit to our spectra plus deconvolved (red) and Morgan data sets between $\lambda 3500 - \lambda 7300$ Å (light blue). The red and light blue shaded areas represent the error in the linear fit associated to the magnitude difference.

To explain the increase of $\Delta m > 0.42$ for wavelength beyond $\lambda 7300$Å, we investigated the possible contamination of the B flux using the data from CASTLES. Fifty percent of flux contamination from the lens galaxy in image B for filter F816W yields $\Delta m = -0.14$. The introduction of galaxy flux produces the opposite effect by moving the $m_B - m_A$ difference in the direction of the lines instead of the continuum. Another alternative is contamination by the quasar host galaxy, considering that HST data show a ring in the bands with anomalous flux.

We used the data between $\lambda 4400 - \lambda 7300$ Å to analyze the chromatic microlensing effect. The differences between the baseline and the continua are in Table 5. As described in section 3, we modeled the system using *Lensmodel* (Keeton 2001) to obtain the parameters to compute the magnification maps. We obtained the convergence and the shear for each image: $\kappa_A = 0.30$, $\gamma_A = 0.29$, $\kappa_B = 1.10$, $\gamma_B = 1.08$. These values are consistent with those obtained by Sluse et al. (2012). We obtained the probability density functions of $r_s$ and p using a logarithmic prior for the size for each data set. As both calculations are independent we performed the product of these results to obtain the probability distribution function in Figure 6. We found that the size and temperature profile of the accretion disk is: $r_s = 3.6^{+3.0}_{-1.6} \sqrt{M/M_\odot}$ light-days ($\ln((r_s/\text{light-days})\sqrt{M/M_\odot}) = 1.3 \pm 0.6$), and p = $2.0 \pm 0.7$ at $1\sigma$. This is the first estimation of the size and p using microlensing for this system. If we compare our results with those expected from the theory we find that the size is larger than the theoretical value $r_{s-theory} = 0.3 \sqrt{M/M_\odot}$ light-days assuming p=4/3, a $M_{BH} = 1.1 \times 10^9 M_\odot$ (Sluse et al. 2012), an Eddington ratio of $L/L_E = 0.1$ and a radiative efficiency of $\eta = 0.1$. The value for p is in agreement, within errors, with the proposed by Shakura & Sunyaev (1973). Considering the results obtained for other systems using this technique (Rojas et al. 2014; Jiménez-Vicente et al. 2015b; Motta et al. 2017), our results confirm that the size of the accretion disk is larger than predicted.

Table 5. Q1355-2257 Chromatic-Microlensing

| Data | $\lambda$ (Å) | $\Delta$m (mag) |
|---|---|---|
| Our + Deconvolved dataset | 4400 | $0.53 \pm 0.09$ |
| | 6200 | $0.39 \pm 0.09$ |
| | 7300 | $0.31 \pm 0.09$ |
| Morgan et al. (2003) | 4400 | $0.68 \pm 0.11$ |
| | 6200 | $0.44 \pm 0.11$ |
| | 7300 | $0.29 \pm 0.11$ |

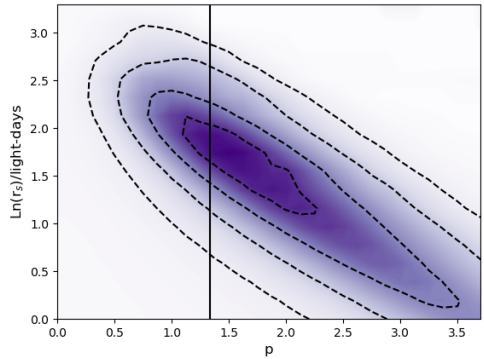

**Figure 6.** Probability density function using a logarithmic size prior for the chromatic microlensing measurements in Q1355-2257. Contours correspond to $0.5\sigma$, $1\sigma$, $1.5\sigma$, and $2\sigma$ respectively. The black solid line shows the value for p by Shakura & Sunyaev (1973).

### 4.3. *SDSS1029+2623*

For this system the source is at $z_s=2.197$. It was discovered by Inada et al. (2006). It was thought to be a double system with a large separation between A and B (22.5"), but Oguri et al. (2008) discovered a third component separated 1.8" from the B component. The large separation between A and B occurs because the lens is a cluster of galaxies at $z_l=0.58$ (Oguri et al. 2008).

We present our own analysis for the spectra in Oguri et al. (2008). The line profiles (Figure 7) show strong absorption in the case of Ly$\alpha$ and CIV. For that reason, we used small windows to integrate the line flux, and in the case of CIV, we split the analysis into two windows (Table 6).

The differences between the core of the emission line and the adjacent continuum are negligible (Figure 8), which is evidence of no microlensing effect in the spectra at that epoch (December 2007). The broad band fluxes presented in Oguri et al. (2008) where taken in November 2006 (B), May 2007 (VRI), and January 2008 (g, and R) respectively. The B band follows the trend of the spectroscopic data, but the rest of the bands show an offset of $\sim 0.2$ mag. The displacement is not attributed to microlensing variation in different epochs because almost all broad band measurements from 2006-2007 are in agreement with those taken in 2008. To investigate whether the emission lines are affecting the broad band measurements, we integrated the spectra in the same wavelength range of the V,R,I,g, and R bands. However, the integration shows the same trend as the measurements in the spectra. A possible explanation is flux loss in the spectra (e.g. $\sim$10% in B component), producing a displacement in the data.

Figure 8 shows that the magnitude differences for both lines and the continua increase towards blue wavelengths, which could be explained by dust extinction produced by a galaxy in the vicinity of the C component (Oguri et al. 2013). The possibility of extinction was previously analyzed by Oguri et al. (2008), obtaining $\Delta E \sim$ 0.15-0.2 using $R_v = 3.1$, and Ota et al. (2012), giving $\Delta E \sim 0.17$ at $z_l = 0.584$.

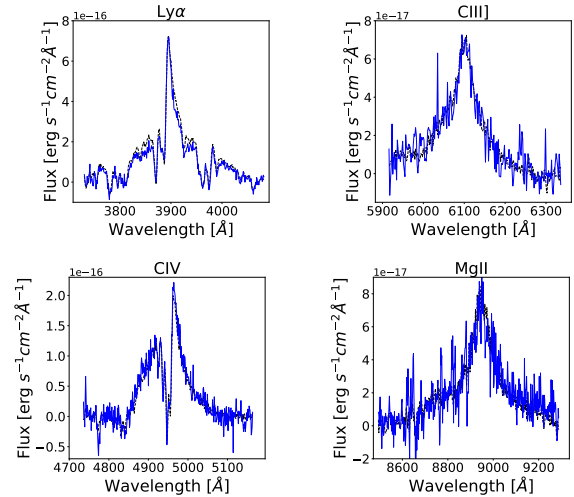

**Figure 7.** Ly $\alpha$, CIII], CIV, and MgII emission lines profiles without the continuum for SDSS1029+2623. In black wre show the A component and in blue the B component multiplied by a factor of 6.0 (Ly $\alpha$ and CIII]), 4.5 (CIV), and 3.3 (MgII].

**Table 6.** SDSS1029+2623 Magnitude Differences

| Region | $\lambda$ (Å) | Window[a] (Å) | $m_C$-$m_B$ (mag) |
|---|---|---|---|
| Continuum | 3905 | 3640-4250 | $2.09 \pm 0.03$ |
| | 4900 | 4700-5200 | $1.84 \pm 0.10$ |
| | 4972 | 4700-5200 | $1.82 \pm 0.10$ |
| | 6103 | 5800-6420 | $1.67 \pm 0.01$ |
| | 8942 | 8300-9417 | $1.35 \pm 0.01$ |
| Line | Ly$\alpha$ | 3890-3920 | $2.03 \pm 0.43$ |
| | CIV (1) | 4878-4923 | $1.86 \pm 0.14$ |
| | CIV (2) | 4960-4985 | $1.84 \pm 0.14$ |
| | CIII] | 6083-6123 | $1.64 \pm 0.01$ |
| | MgII | 8925-8960 | $1.40 \pm 0.02$ |

[a] integration window.

We used the spectroscopic data to perform a new extinction analysis under the assumptions that the absorber is the galaxy in the vicinity of component C. Considering the absence of microlensing, we combine both the emission line and the continuum magnitude differences to fit an extinction curve at the redshift of the lens. We present two cases, for the first one we left $R_v$ as a free parameter and for the second one we fixed $R_V = 3.1$, which corresponds to the average value for our galaxy. The best fit parameters for the first case are: $\Delta M_0 = 1.16 \pm 0.01$, $\Delta E = 0.22 \pm 0.01$, and $R_V = 3.24 \pm 0.02$ ($\chi^2 = 39.93$, degree of freedom = 11). For the second case the best fit parameters are: $\Delta M_0 = 1.20 \pm 0.01$, and $\Delta E = 0.20 \pm 0.01$ ($\chi^2 = 205.5$, degree of freedom = 12). In both cases we obtain similar results for the extinction in agreement with previous results by Oguri et al. (2008) and Ota et al. (2012).

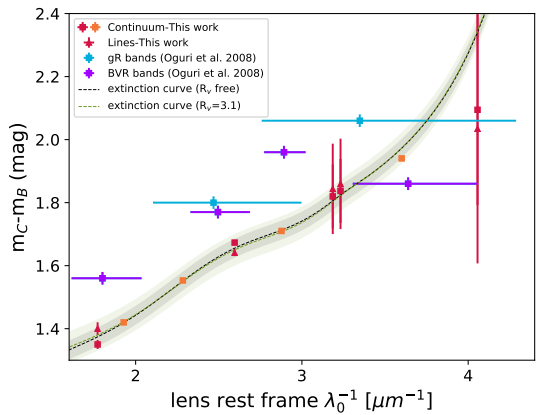

**Figure 8.** Magnitude differences $m_C - m_B$ as a function of wavelength for SDSS1029+2623. The wavelength axis is plot in the lens rest frame for a better interpretation of the extinction curve. The triangles are the values for the emission line cores without continuum and the squares are the continuum integration from the spectra or band, depending on the data set. Red and orange symbols are measurements from the spectra. Data from Oguri et al. (2008) are plotted in light blue (bands g and R, taken in 2008 with the Keck telescope) and in purple (bands, B, V, R, and I, taken in 2006-2007 with the UH88 telescope). The black (green) dashed line is the extinction curve fitted to the spectroscopic data using all the parameters free (fixing $R_v = 3.1$), the grey (light green) shaded region represents the error associated with the magnitude difference.

## 5. CONCLUSIONS

We used spectroscopic data of the lensed quasars: SDSS0924+0219 (BC), Q1355-2257 (AB), and SDSS1029+2623 (BC) to study their flux anomalies. We obtained the following results:

1. Comparing the magnitude differences of the cores of the emission lines with those of the continua we found chromatic microlensing in SDSS0924+0219, and Q1355-2257. We estimate the accretion disk at $\lambda_{rest-frame} = 3533$Å size and temperature profile. Although the CIII] and MgII line profiles of SDSS0924+0219 and Q1355-2257 present slight differences in the wings, only those in the red wing of CIII] in SDSS0924+0219 are large enough to be likely due to microlensing in the BLR, but only one epoch is not enough to constrain the size of this region. In the case of SDSS1029+2623 we found evidence of extinction, produced by a galaxy near C and no evidence of microlensing.

2. In the case of SDSS0924+0219 we obtain $r_{s,\lambda 3533} = 4^{+3}_{-2} \sqrt{M/M_\odot}$ light-days and $p = 0.8 \pm 0.2$. The size is in agreement with those of Floyd et al. (2009) and Blackburne et al. (2011). The estimation for p is significantly smaller than the value expected from the theory, but in agreement with the average value found by Jiménez-Vicente et al. (2014) in a sample of 8 lensed quasars.

3. Q1355-2257 shows chromatic microlensing which allows us to estimate $r_{s,\lambda 3533} = 3.6^{+3.0}_{-1.6} \sqrt{M/M_\odot}$ light-days and $p = 2.0 \pm 0.7$. This is the first estimate of the accretion disk size and temperature profile for this system. Comparing with theory we found that the size is larger than, and p is in agreement within errors with predictions.

4. For SDSS1029+2623 we fitted an extinction curve to the data assuming that the galaxy acting as absorber is at the redshift of the cluster. We study two cases, one with $R_V$ as a free parameter and other where we fixed $R_V = 3.1$. In both cases we obtained a value for the extinction around $\Delta E = 0.2$ that is in agreement with previous estimations (Oguri et al. 2008; Ota et al. 2012).

The authors are grateful to M. Oguri for kindly providing the spectra of SDSS1029+2623. K.R. acknowledge support from PhD fellowship FIB-UV 2015/2016, Becas de Doctorado Nacional CONICYT 2017, LSSTC Data Science Fellowship Program, her time as a Fellow has benefited this work, and support from the Swiss National Science Foundation (SNSF). V.M. acknowledge partial support from Centro de Astrofísica de Valparaíso. C.F. acknowledge support of La Caixa Fellowship. J.J.V. is supported by project AYA2017-84897-P financed by the Spanish Ministerio de Economía y Competividad and by the Fondo Europeo de Desarrollo Regional (FEDER), and by project FQM-108 financed by Junta de Andalucía.

*Facility:* VLT/FORS, HST

*Software:* IRAF, Python: Numpy, Matplotlib, Scipy

## APPENDIX

## A. SPECTROSCOPIC DATA

We present the figures of the spectroscopic data sets described in Section 2 (Figures 9, 10, 12) . Also we include the figure of the deconvolved spectra of Q1355-2257 presented in Sluse et al. (2012) that we reanalyzed (Figure 11).

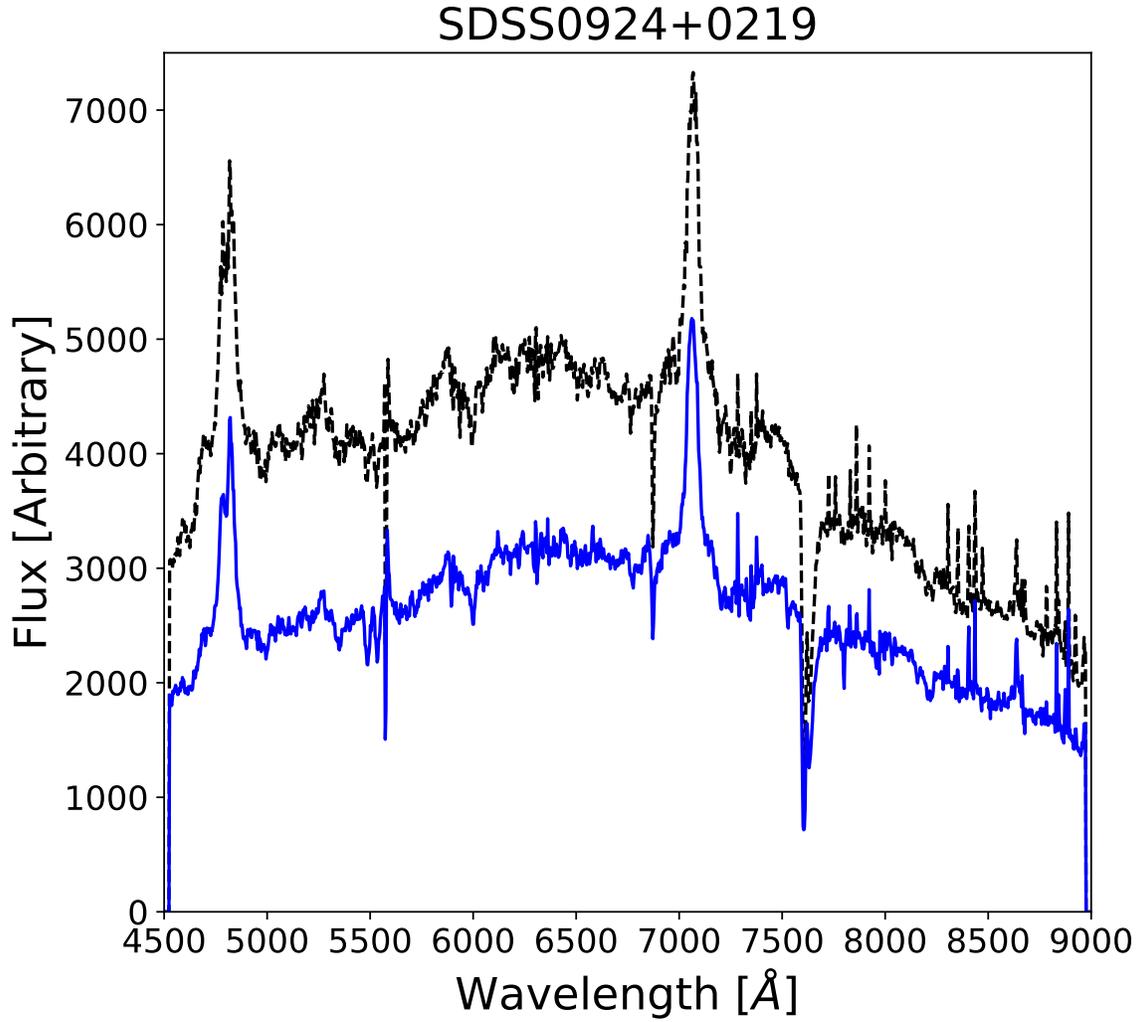

**Figure 9.** Spectra from VLT/FORS2 observations taken in 2008 for SDSS0924+0219. In black the A component and in blue the B component.

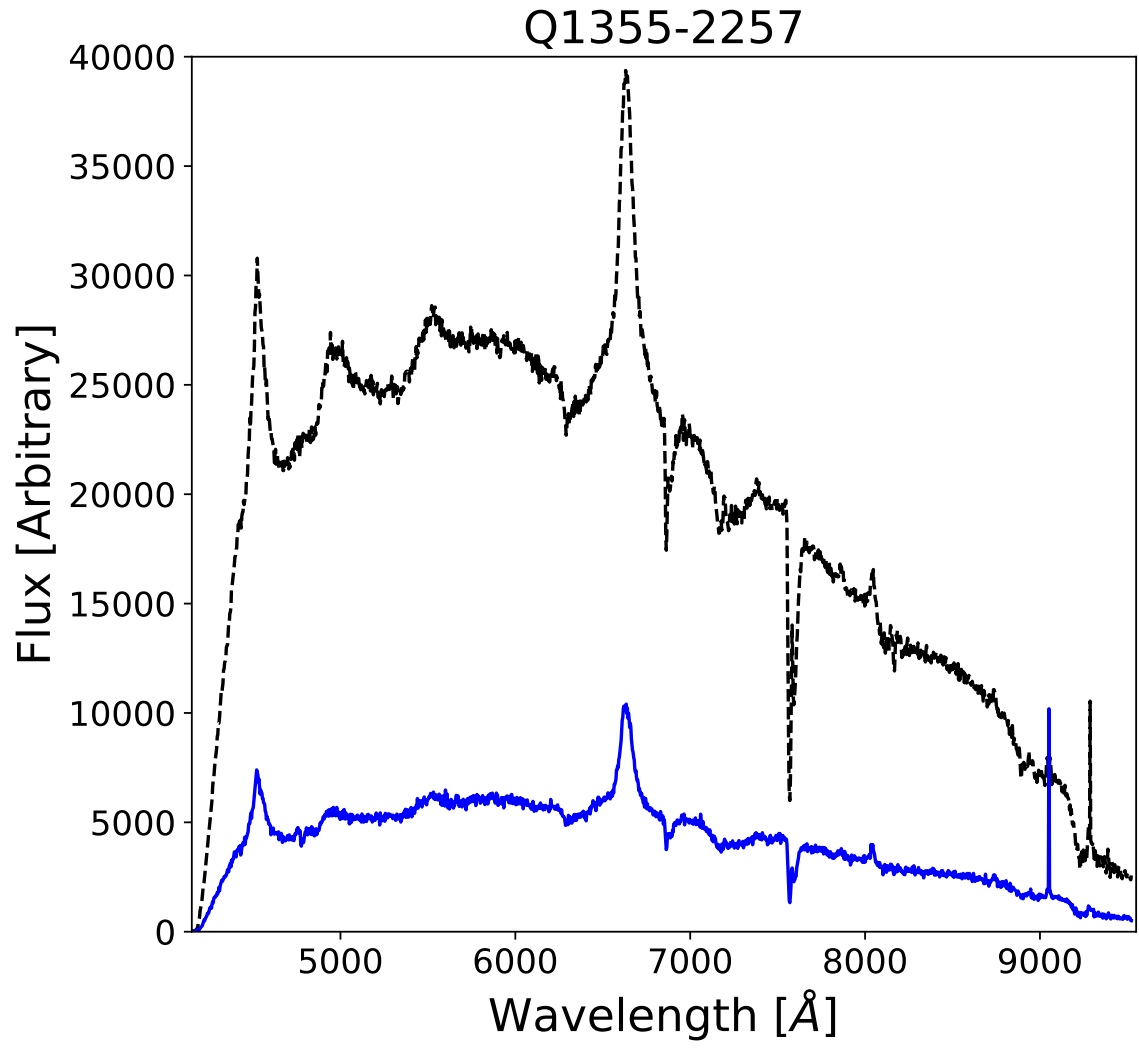

**Figure 10.** Spectra from VLT/FORS2 observations taken in 2008 for Q1355-2257. In black the A component and in blue the B component.

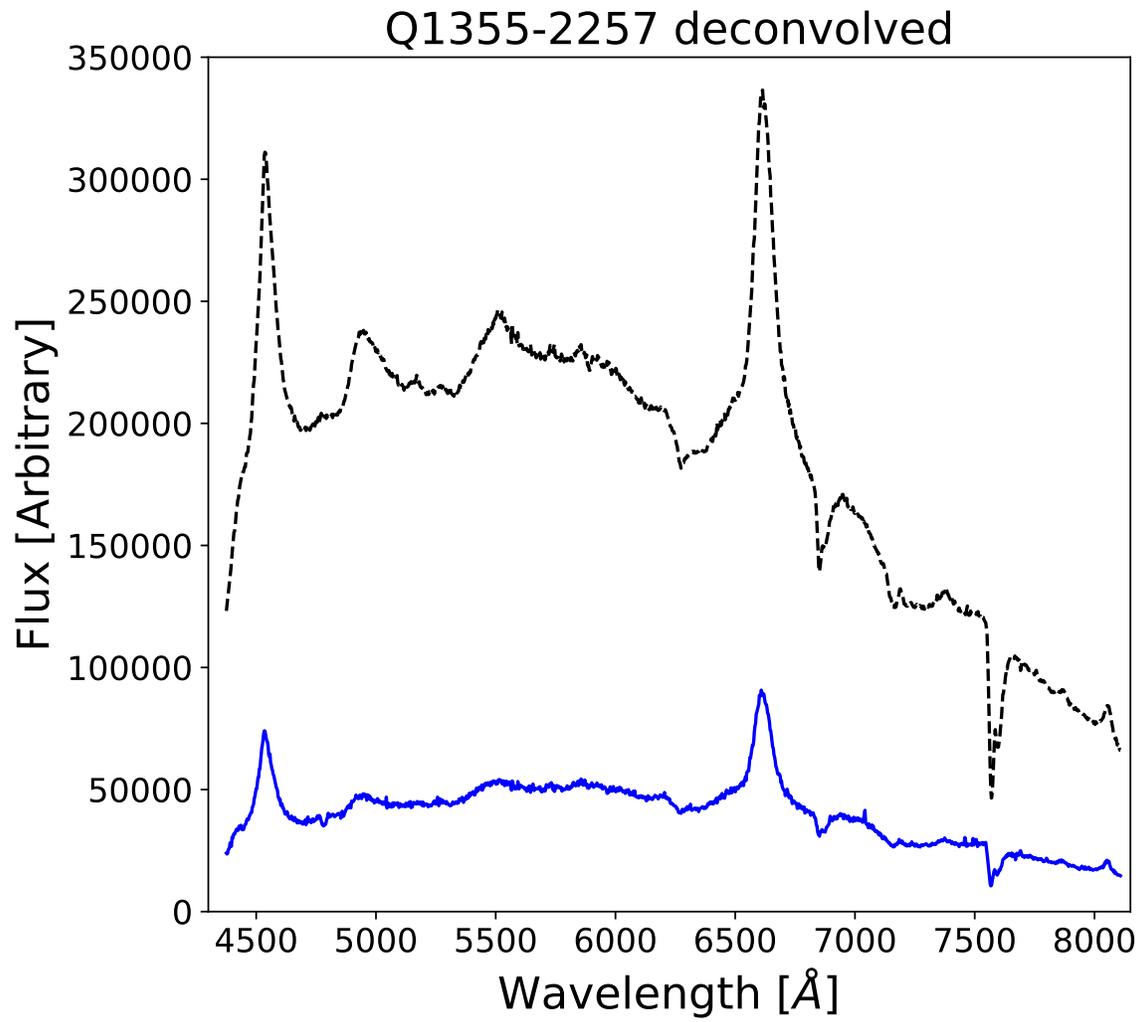

**Figure 11.** Deconvolved spectra for Q1355-2257 presented in Sluse et al. (2012). The data was observed with VLT/FORS1 in 2005. In black the A component and in blue the B component.

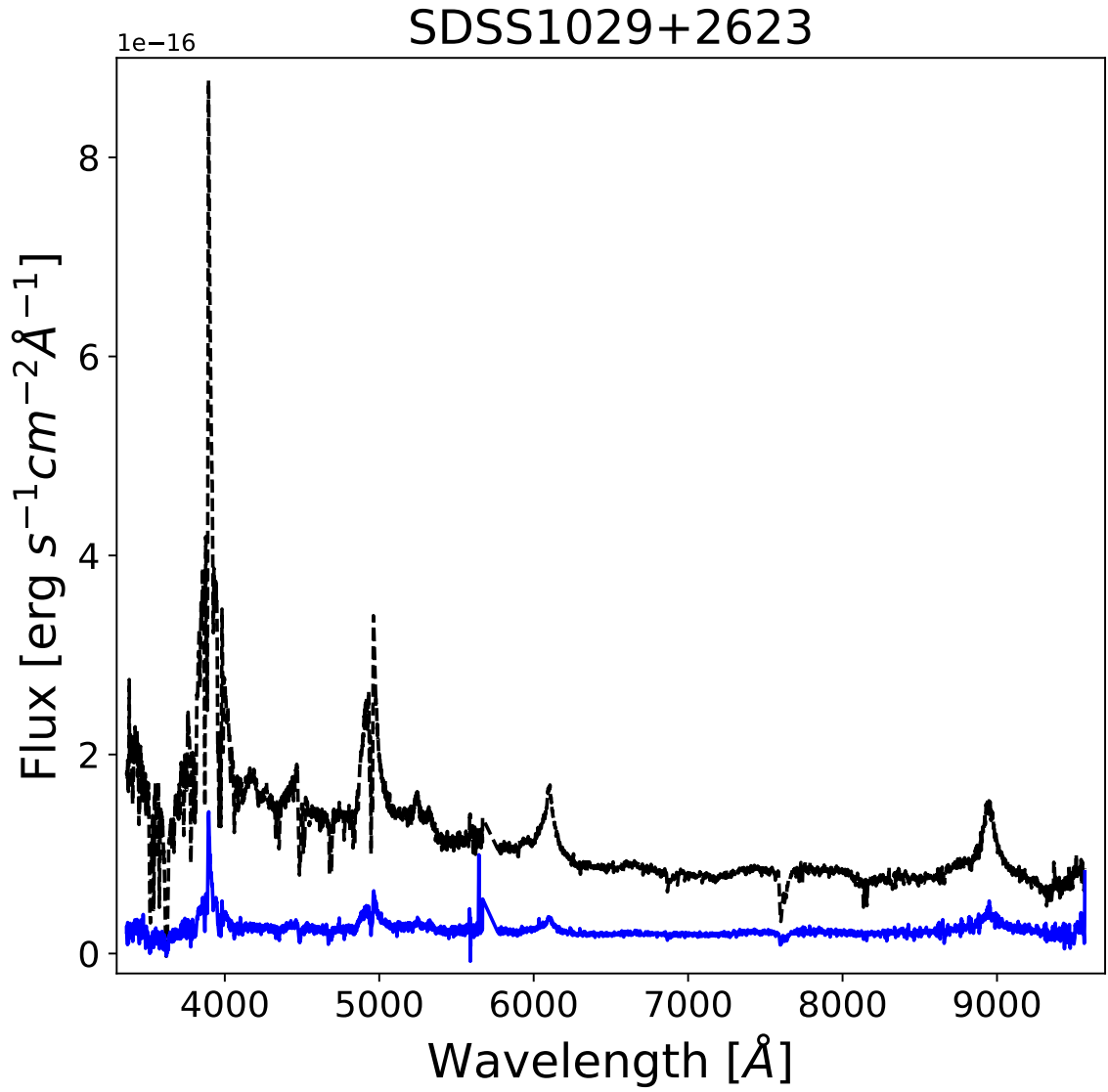

**Figure 12.** Spectra from LRIS-ADC at Keck telescope for SDSS J1029+2623 provided by M. Oguri.